# FQPDR: Federated Quantum Neural Network for Privacy-preserving Early Detection of Diabetic Retinopathy

**Debashis De · Mahua Nandy Pal · Dipankar Hazra**

**Abstract** Diabetic Retinopathy (DR) is a common complication of diabetes that can lead to blindness of people. Detecting DR at the earliest stage is essential to prevent irreversible eye damage. Microaneurysm dots are the first signs of DR. As the dots are tiny and of low contrast, detecting mild DR is a very challenging task. Federated learning (FL) preserves data privacy, which is a major concern for medical image processing. FL is a collaborative learning method, which shares only the model parameters with a server, without sharing the patient data to a central server. Inspired by classical FL, we propose a federated learning-based quantum neural network (federated QNN) for this task. We implemented the models with limited samples and few learnable parameters from the E-ophtha and Retina MNIST datasets. The cross-evaluation efficiency of the proposed federated quantum neural network system for privacy-preserving early detection of diabetic retinopathy (FQPDR) in Kaggle dataset images indicates the robustness of the light weight learning

Debashis De
Department of Computer Science and Engineering, Maulana Abul Kalam Azad University of Technology, West Bengal
E-mail: dr.debashis.de@ieee.org

Mahua Nandy Pal
Department of Computer Science and Engineering, MCKV Institute of Engineering, West Bengal
E-mail: mahua.nandy@gmail.com

Dipankar Hazra
Department of Computer Science and Engineering, OmDayal Group of Institutions, West Bengal
E-mail: dipankar1998@rediffmail.com

models. FQPDR performances are inspiring while considering existing non-FL and FL methods.

## 1 Introduction

Diabetic retinopathy (DR) is one of the leading causes of blindness in diabetic patients.

Early detection of diabetic retinopathy is crucial to prevent irreversible changes in eye. Detecting DR at the earliest stage from an image of the retinal fundus is a challenging task due to the low contrast and tiny characteristics of DR symptoms in earlier stages. Microaneurysm dots are the earliest indicators of DR. Detecting mild DR is particularly challenging due to the subtle nature of these dots. The objective of this work is the early diagnosis of DR by identifying mildly affected images.

There are works on the early detection of diabetic retinopathy using convolution neural network (CNN) which achieves good accuracy. The drawbacks of these implementations are the requirement of an ever-growing number of samples, high memory and time complexity for effective network learning and large number of learnable parameters. According to some notable developments [1],[2], QNN has potentially better pattern generalization capability. In a quantum neural network (QNN), the cost function and gradients are computed using a parameterized quantum circuit (PQC). The quantum circuit uses a smaller number of parameters than a classical CNN to achieve the same accuracy. In the Noisy Intermediate Scale Quantum (NISQ) era, shallow-depth quantum circuits can be implemented with fewer qubits for PQC. In this paper, we propose a federated QNN-based system for DR-affected image identification at an early stage (FQPDR). We implemented an amplitude encoding method to convert image patches to qubit information that will be taken as input for the quantum neural network (QNN). Gradient Descent, Nesterov Momentum and Adam optimizers are evaluated to optimize the QNN parameters. Adam Optimizer is best suited for the purpose.

One of the major concerns of medical image processing is to preserve the privacy of the patient data. Thus, it is essential to preserve the privacy and security of patient data. Federated training of quantum model ensures privacy preservation through distributed learning. We implemented federated QNN and received satisfactory results on different edges with local data. After local training with limited data, the model parameters are shared with the federated server. The Federated server computes global model parameters from different local models and shares the parameters of the global model with individual edges, where further training epochs are executed on the global model till desired convergence. The process is repeated several times to achieve better accuracy.

To facilitate smart healthcare systems empowered by the Internet of Medical Things (IoMT) in remote areas, retinal fundus images would be captured with special purpose cameras installed on edge devices. So, FQPDR has the potential to direct towards secure, light weight and intelligent IoMT application.

## 1.1 Motivation

Quantum Neural Networks (QNNs) offer potential advantages over classical Neural Networks (NNs) due to the inherent properties of quantum superposition and entanglement. Superposition enables a qubit to exist in multiple states simultaneously, providing parallelism in computation. The quantum feature space can capture intricate patterns and correlations in the data, leading to the generation of powerful and expressive models with a limited number of samples and limited learnable parameters. Learning a few parameters makes the training process less computationally intensive and faster compared to classical approaches. We implement QNN architecture for DR detection to reap these advantages.

The research is inspired to observe not only the performance of the QNN model, but also the QNN performance in a privacy-preserving architecture of distributed federated learning. Centralized learning approach allows for model training with potential accuracy. However, this approach consumes powerful computational resources to process a large number of training data. In addition, centralized machine learning concerns data privacy and security. Collecting sensitive data in a central repository poses risks of data breaches and unauthorized access. In addition, transferring large volumes of data to a central repository is inefficient and costly in terms of bandwidth and storage. Federated learning allows the training of machine learning models across multiple decentralized devices without the need to transfer sensitive patient data to a central server. This decentralized approach helps to maintain the confidentiality and security of medical data. Additionally, FL can be more scalable and efficient, as it reduces the need for large-scale data transfer and leverages the computational power of edge nodes.

## 1.2 Contribution

The contributions of the article are as follows:

- A QNN based system is implemented for early DR detection involving three retinal image datasets: E-ophtha, Kaggle and MNIST datasets with images of highly diverse resolutions.
- Federated learning is implemented to ensure enriched and diversified model experience and medical data privacy.
- Few parameter QNN is trained with limited samples to learn the feature correlation, as acquiring sufficient annotated medical data is tedious.

– Finally a performance analysis is demonstrated with existing classical and quantum FL/Non-FL methods.

### 1.3 Paper Organization

Section 2 describes the preliminary concepts of quantum computing, section 3 is the literature survey. Literature survey is represented in tabular form in 4 relevant sub-domains - survey on non-federated learning, applied to retinal images (classical methods), survey on federated learning, applied to retinal images (classical methods), survey on different QML approaches, applied to medical images and survey on different FL approaches, applied to other medical images (classical methods). Section 4 describes the proposed FQPDR system. This section encompasses image pre-processing, quantum state preparation, quantum neural network, federated learning and quantum neural network within federated learning architecture. Section 5 presents implementation details, experimental results and qualitative evaluation of the system. This section also compares FQPDR with other existing DR detection methods and with a limited parameter classical CNN system. Section 6 is the discussion about the proposed light-weight federated QNN model-based DR detection system. Section 7 mentions the potential limitations of the system. Section 8 concludes the paper.

## 2 Quantum Computing : Preliminary Concepts

Some fundamental concepts are discussed in this section, which will facilitate the understanding of a parameterized quantum circuit and its optimization, discussed in the following subsections 4.3 and 5.3. Quantum computing is built upon the principles of quantum mechanics, where polarization of photons or spin of the electron decides the mathematical notation of states. Horizontal polarization of the photon or the up spin of the electron denotes the basis quantum state $|0\rangle$ and vertical polarization of photon or down spin of the electron denotes the basis state $|1\rangle$. A qubit can also exists in a linear combination of the basis states $|0\rangle$ and $|1\rangle$, which is called superposition. The two basis states $|0\rangle$ or $|1\rangle$ can be represented in equation (1) and equation (2) respectively.

$$|0\rangle = \begin{bmatrix} 1 \\ 0 \end{bmatrix} \tag{1}$$

$$|1\rangle = \begin{bmatrix} 0 \\ 1 \end{bmatrix} \tag{2}$$

The superposition state can be represented as an equation (3).

$$|\psi\rangle = \alpha\,|0\rangle + \beta\,|1\rangle \tag{3}$$

Where $\alpha$ and $\beta$ are complex numbers and $|\alpha|^2 + |\beta|^2 = 1$ due to normalization condition.

Some basic quantum gates such as Pauli-X or bit flip gate , Pauli-Y or bit and phase flip gate, and Pauli-Z or phase flip gate can be represented by equation (4), equation (5) and equation (6) which works on qubits.

$$PauliX = \begin{bmatrix} 0 & 1 \\ 1 & 0 \end{bmatrix} \tag{4}$$

$$PauliY = \begin{bmatrix} 0 & -i \\ i & 0 \end{bmatrix} \tag{5}$$

$$PauliZ = \begin{bmatrix} 1 & 0 \\ 0 & -1 \end{bmatrix} \tag{6}$$

Rx, Ry, and Rz gates are basic parameterized gates. These are fundamental single-qubit rotation gates. Rx, Ry and Rz gates are parameterized by an angle ($\theta$) and rotates the qubit state by that angle around the X, Y and Z-axes respectively. The matrix representations of Rx, Ry and Rz gates are shown in equations (7), (8) and (9) respectively.

$$Rx(\theta) = \begin{bmatrix} \cos(\frac{\theta}{2}) & -i\sin(\frac{\theta}{2}) \\ -i\sin(\frac{\theta}{2}) & \cos(\frac{\theta}{2}) \end{bmatrix} \tag{7}$$

$$Ry(\theta) = \begin{bmatrix} \cos(\frac{\theta}{2}) & -\sin(\frac{\theta}{2}) \\ \sin(\frac{\theta}{2}) & \cos(\frac{\theta}{2}) \end{bmatrix} \tag{8}$$

$$Rz(\theta) = \begin{bmatrix} e^{-i\frac{\theta}{2}} & 0 \\ 0 & e^{i\frac{\theta}{2}} \end{bmatrix} \tag{9}$$

Another important property of quantum computing is called entanglement which enables qubits to be correlated in such a way that classical bits are unable to do. For example, the Controlled-X or CNOT gate is an entanglement gate. It works on two qubits. The CNOT gate flips the target qubit $q_1$ if the control qubit $q_0$ is in $|1\rangle$ state. The SWAP gate exchanges the states of two qubits. Similar to CNOT gate, the Toffoli gate is a three-qubit gate with two control qubits and one target qubit. It flips the state of the target qubit if and only if both control qubits are in the state $|1\rangle$. The entanglement gates CNOT, Swap and Toffoli gate are shown in equations (10), (11), and (12) respectively.

$$\text{CNOT}(q_0,q_1) = \begin{bmatrix} 1 & 0 & 0 & 0 \\ 0 & 1 & 0 & 0 \\ 0 & 0 & 0 & 1 \\ 0 & 0 & 1 & 0 \end{bmatrix} \tag{10}$$

$$\text{SWAP}(q_0,q_1) = \begin{bmatrix} 1 & 0 & 0 & 0 \\ 0 & 0 & 1 & 0 \\ 0 & 1 & 0 & 0 \\ 0 & 0 & 0 & 1 \end{bmatrix} \tag{11}$$

$$\text{Toffoli}(q_0,q_1,q_2) = \begin{bmatrix} 1 & 0 & 0 & 0 & 0 & 0 & 0 & 0 \\ 0 & 1 & 0 & 0 & 0 & 0 & 0 & 0 \\ 0 & 0 & 1 & 0 & 0 & 0 & 0 & 0 \\ 0 & 0 & 0 & 1 & 0 & 0 & 0 & 0 \\ 0 & 0 & 0 & 0 & 1 & 0 & 0 & 0 \\ 0 & 0 & 0 & 0 & 0 & 1 & 0 & 0 \\ 0 & 0 & 0 & 0 & 0 & 0 & 0 & 1 \\ 0 & 0 & 0 & 0 & 0 & 0 & 1 & 0 \end{bmatrix} \tag{12}$$

Quantum computers can exploit these two quantum phenomena, superposition and entanglement, to perform certain calculations efficiently in an exponentially faster way than classical computers, to achieve "Quantum Supremacy".

## 3 Literature Survey

A literature survey of the proposed method is concisely discussed in tabular form. Table 1 shows a survey on classical non-federated learning, applied to retinal images, whereas Table 2 describes a survey on classical federated learning, applied to retinal images. Table 3 displays a survey on QML approaches, applied to medical images. Table 4 presents a survey of classical federated learning approaches, applied to other medical images. Besides these references, some of the existing significant algorithms for data privacy preservation are available in [4], [5] etc. and some classical disease prediction works are [6], [7], and [8] etc.

## 4 Proposed FQPDR for Privacy-preserving Early Detection of DR

The proposed Federated QNN based FQPDR system consists of the following steps: image pre-processing, quantum state preparation, application of quantum neural network within federated learning architecture. The block diagram of the proposed system is shown in figure 1. We discuss these steps in the following subsections.

Table 1: Survey on Non-federated Learning, Applied to Retinal Images (Classical Methods)

| Author-year | Title | Proposed Method | Limitation |
|---|---|---|---|
| Kumar B. N., Mahesh T. R., Geetha G., Guluwadi S., -2023 | Redefining Retinal Lesion Segmentation: A Quantum Leap With DL-UNet Enhanced Auto Encoder-Decoder for Fundus Image Analysis.[9] | Deep learning based affected region segmentation method. Step 1: Patches are extracted of 512x512 for original and ground truth images. Step 2: These patches are fed as input of a U-Net. Step 3: Channel-wise spatial attention mechanism refine the MA segmentation task. | The method is not checked for spatial relationship between pixels, input variations and class imbalance problems. |
| Tavakoli M.,Mehdizadeh A., Aghayan A., Sahari R.P., Ellis T., Dehmeshki J., -2021 | Automated Micro aneurysms Detection in Retinal Images using Radon Transform and Supervised Learning: Application to Mass Screening of Diabetic Retinopathy.[11] | Affected region segmentation method. Step 1: Background variations are removed. Step 2: Optic nerve head and retinal vessels are masked using radon transform and multi-overlapping windows. Step 3: MA segmentation using a combination of Radon Transform and Supervised Support Vector Machine. Specificity = 95.7%, sensitivity = 95.46%, AUC=97% | High average processing time, which is 4.5 minutes per image. |
| Shorfuzzaman M, Hossain MS, El Saddik A. -2021. | An Explainable Deep Learning Ensemble Model for Robust Diagnosis of Diabetic Retinopathy Grading.[13] | Affected retinal image identification method. Step 1: Deep learning ensemble model where weights from different models are fused into a single model. Step 2: Extracts salient features from various retinal lesions found on fundus images. Precision =97%, Sensitivity = 98%, AUC=97.8%.for APTOP, Messidor and IDRiD dataset. | Large number of samples and learnable parameters |

Table 2: Survey on Federated Learning, Applied to Retinal Images (Classical Methods)

| Author-year | Title | Proposed Method | Limitation |
|---|---|---|---|
| Soni M., Singh M.K., Das P., Shabaz M., Shukla P. K., Sarkar P., Singh S., Keshta I., Rizwan A., -2023 | IoT-Based Federated Learning Model for Hypertensive Retinopathy Lesions Classification. [14] | Local features of Arterial and Venous Nicking (AVN) are fused with global features using mean. Step 1: The AVN model sends the image directly to SeqNet for blood vessel segmentation, and arterial and venous classification and uses the intersection method to detect all arteries and veins in the image. Step 2: ResNet-50 is used for classification. Step 3: The IoT-FHR model directly sends the pre-processed fundus images to the VGG-19 for classification. Step 4: Finally, the weighted average of the two features is sent to a simple neural network for classification. The accuracy, sensitivity and specificity of this model are 93.50%, 69.83% and 98.33% respectively. | An advanced model fusion algorithm is needed to increase the model's sensitivity. Computation exhaustive method involving a number of learning models. |
| Jagan Mohan N., Murugan R., Goel T., Roy P., -2023 | DRFL: Federated Learning in Diabetic Retinopathy Grading Using Fundus Images. [18] | DR detection and grading method. The central server is trained with a client database with randomly initialized weight. In the next steps database from the second client is trained with updated weight from a central server and the weight is updated again in a central server. Then third client database is trained with updated weight sent by the central server and this process continues. The proposed model achieves an accuracy of 98.6%, specificity of 99.3%, precision of 97.25% and F1-score of 97.5%. | Large number of samples and learnable parameters. |

Table 3: Survey on QML Approaches, Applied to Medical Images

| Author-year | Title | Proposed Method | Limitation |
|---|---|---|---|
| Chen, G., Chen, Q., Long, S., Zhu, W., Yuan, Z. and Wu, Y.-2022. | Quantum Convolutional Neural Network for Image Classification. [20] | The image data is downscaled with multi-scale entanglement renormalization ansatz (MERA) and box counting-based fractal scale features. The 10 qubit input features are fed into QCNN. Quantum convolution and quantum pooling layers are constructed by a cascade of two-qubit parameterized unitary. One QCNN and two hybrid quantum-classical models are compared with the classical model with the breast cancer dataset. The method has accuracies up to 98%. | More different types of images and other feature extraction methods to be considered. |
| Choudhuri, R. and Halder, A.-2023. | Brain MRI Tumour Classification using Quantum Classical Convolutional Neural Net Architecture. [21] | In this hybrid model, the quantum convolution layer (QCL) works in the first 3 steps and the classical convolution layer (CCL) works in the next 3 steps as follows: Step 1: Image pixel intensities are angle encoded using RX, RY and RZ gates. Step 2: A series of unitary transformations finds extract features from images. Step 3: The data is decoded or measured. The output of QCL goes to CCL. Step 4: Classical convolution layers detect patterns in images. Step 5: Max-pool layer used for subsampling image data. Step 6: A fully Connected Layer with a sigmoid function is used for binary classification. The proposed method has accuracy on Brats dataset =98.72%, Harvard dataset =98.46%, and private dataset =98.17%. | The extraction ability of QCL can be improved. Multi-class image classification is not tested. |
| Kulkarni, V., Pawale, S. and Kharat, A., 2023. | A Classical–Quantum Convolutional Neural Network to Detect Pneumonia from Chest Radiography.[23] | Step 1: The images are resized to 128x128 and normalized by dividing by 255. So, all pixel values become 0 to 1. Step 2: The classical neural network has a total of 11 layers; ReLU is used activation function in all intermediate layers, sigmoid function is the activation function of the final layer. Step 3: Vanilla gradient descent optimization is used with a constant learning rate of 0.01. Step 4: The penultimate layer is replaced by VQC. Accuracy for detecting pneumonia on chest radiograph = 74. 6%. | Performance needs to be checked in a quantum processor. |
| S.S.Kavitha, Narasimha Kaulgud, 2022. | Quantum machine learning for support vector machine classification.[24] | This method compares the execution speed and accuracy of the quantum support vector machines with classical support vector machines. Proper feature map selection for complex benchmark data sets is most important. | More dataset evaluation is required. |

Table 4: Survey on Federated Learning, Applied to Other Medical Images (Classical Methods)

| Author-year | Title | Proposed Method | Limitation |
|---|---|---|---|
| Kareem A., Liu H., Vladan Velisavljevic V.,-2023 | A Federated Learning Framework for Pneumonia Image Detection using Distributed Data. [15] | In this FL framework, data is equally distributed and trained at 4 different virtual devices separately. After completion of the local training, the models are sent to the aggregated server. In the server, a central model is created and this central is sent back to the local devices. The process is repeated using model parameters. The accuracy, precision, recall, F1-score of the model are 95%, 97%, 96%, 96% respectively. | Quality of data, communication overhead, and healthcare regulation are the main limitations of the method. |
| Islam M., Reza T., Kaosar M., Parvez M. Z., -2022 | Effectiveness of Federated Learning and CNN Ensemble Architectures for Identifying Brain Tumors using MRI Images. [16] | This FL based method identifies brain tumours from MRI images, and consists of the following steps: Step 1: 6 different pre-trained CNN models test the dataset on local devices. Step 2: 3 best CNN models chosen based on accuracy. Step 3: Construct a global model at the server based on the voting ensemble and send it to local devices. | Tiny dataset and poor-quality images are the limitations of this method. |

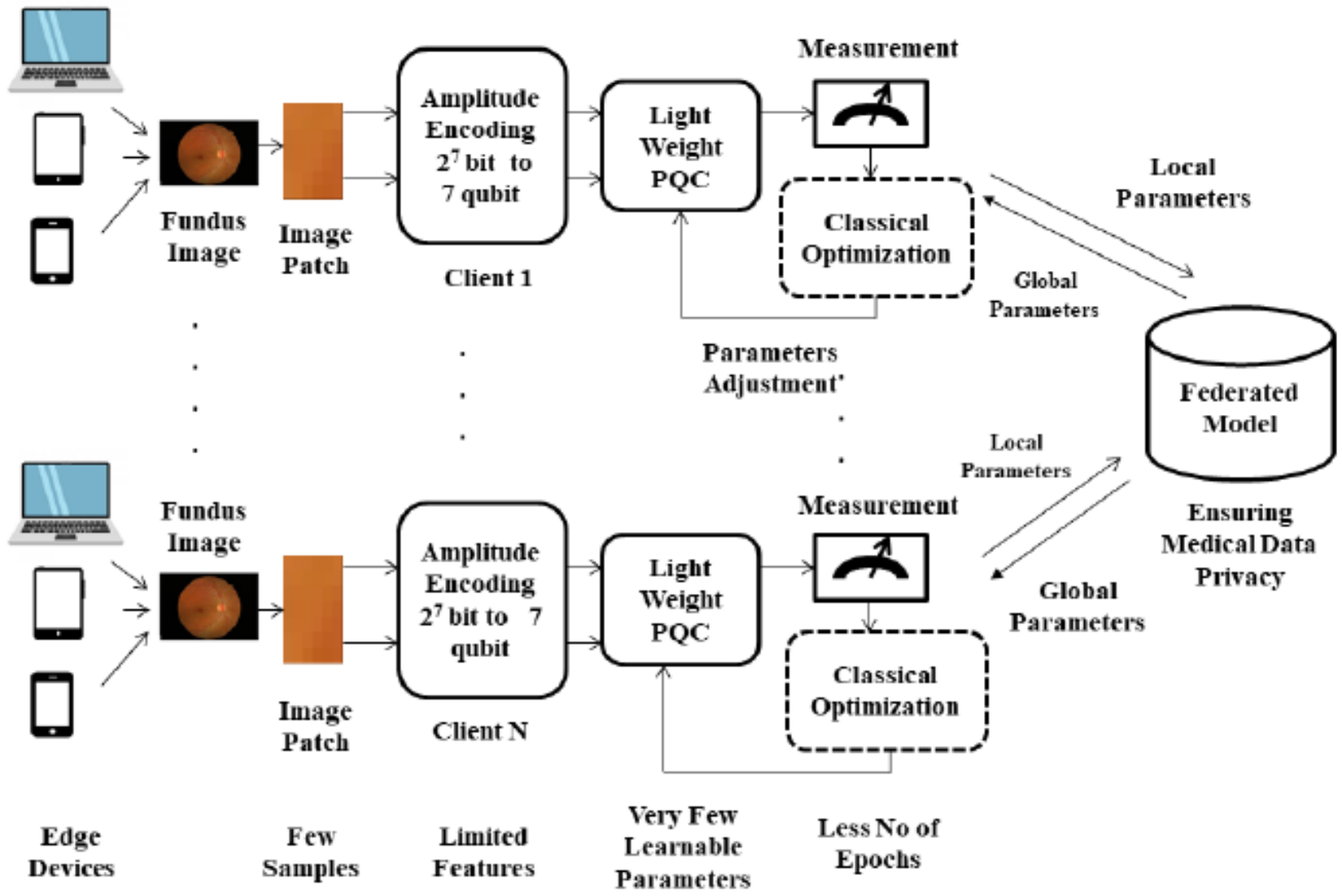

Fig. 1: Block Diagram of the Light-weight Federated QNN Model based DR Detection System

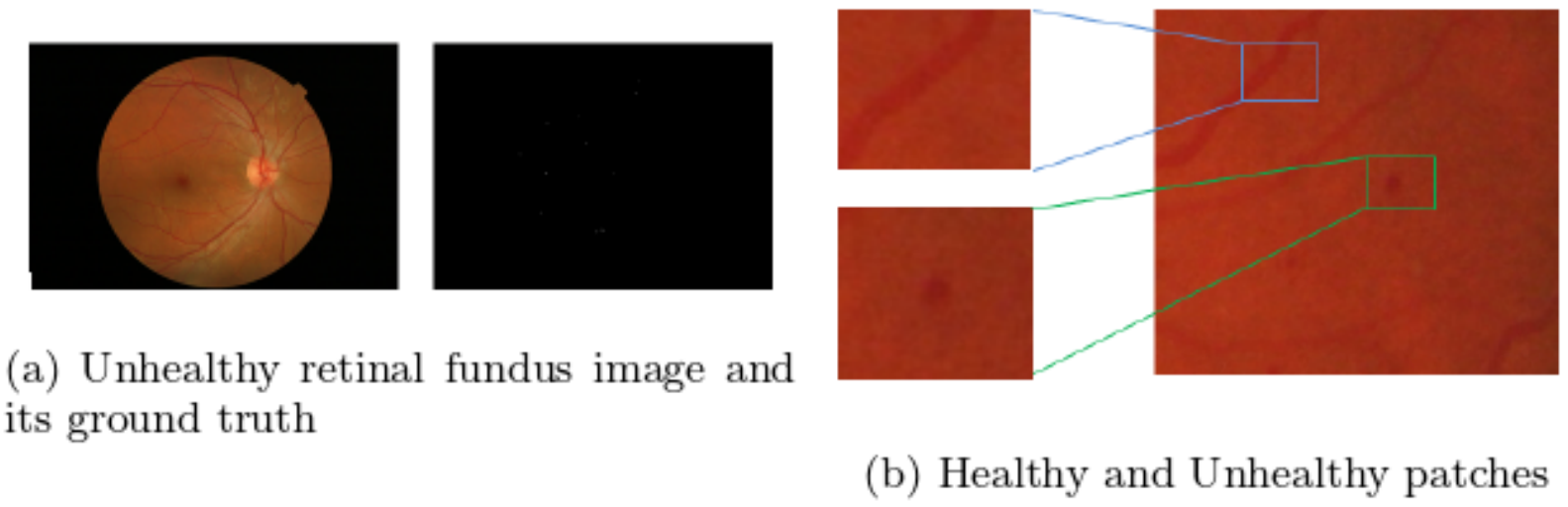

(a) Unhealthy retinal fundus image and its ground truth

(b) Healthy and Unhealthy patches

Fig. 2: Retinal image patches and corresponding annotations

### 4.1 Image Pre-processing

An early-stage DR-affected image and its annotation by the specialist are shown in figure 2a. We have extracted the same number of mild DR MA affected patches and healthy patches for training and validation by a quantum neural network. The patch size is chosen as 7x6x3. An example healthy patch is shown as the blue highlighted region in figure 2b, whereas a patch with mild DR symptom is shown in the same figure by the green highlighted region.

### 4.2 Quantum State Preparation

In the quantum state preparation step, image information is converted to qubits to be fed into a parameterized quantum circuit (PQC).

We applied amplitude encoding technique for representing classical information into quantum states. The image patch resolution is 7x6x3=126 with two extra zero bits padded before applying the encoding operation. Amplitude encoding can convert $2^n$ bit to n qubit or in other words, it converts a vector of length n into $\log_2(n)$ qubit quantum state. The input needs to satisfy the normalization condition $|x|^2$ =1. We have taken $128 = 2^7$ bit data as input. Thus, We have used a 7-qubit QNN for our system implementation. The state vector of the quantum state $\psi(x)$ prepared by amplitude encoding is represented by the equation (13).

$$|\psi(x)\rangle = \sum_{i=1}^{n} x_i \, |i\rangle \tag{13}$$

where $|i\rangle$ is the computational basis for the Hilbert space, $x_i$ is the i-th term of classical feature vector x.

### 4.3 Quantum Neural Network

QNN differs from classical NN in the context of detecting diabetic retinopathy from images of the retinal fundus. QNN encodes classical data into quantum states. Quantum phenomena, superposition, and entanglement promote the potential representation of complex data. Quantum algorithms can perform certain computations exponentially faster than classical algorithms. For diabetic retinopathy detection, this means QNN could analyze large data and intricate patterns in retinal images efficiently, improving diagnostic accuracy and speed. QNNs scale better for high-dimensional data, offering efficient processing of medically significant image data. QNN and classical neural network (NN) follow the same sequence of execution steps for training by optimizing the network parameters to find a favorable result. But the basic difference in execution steps is as follows. In classical NN, parameters are typically weights and biases, associated with neurons and the operations are linear transformations followed by non-linear activation functions. In quantum neural network, parameters are associated with quantum gates in the circuit and operations are quantum gate applications. In both cases, gradients of the loss are computed with respect to the parameters and parameters are updated using a classical optimization algorithm.

A QNN model comprises three primary phases: quantum data encoding or quantum state preparation, quantum circuit generation and measurement. The block diagram illustrating a QNN model is shown in figure 3. A parameterized quantum circuit (PQC) comprises one or more layers of parametric and non-parametric quantum gates. The quantum gates perform rotation and entanglement operations on qubits. In quantum entanglement the characteristics of two or more qubits become linked such that the condition of one affects the condition of the other. This enforces the capability of QNN to capture information correlation effectively. A PQC, also referred to as a variational quantum circuit (VQC), relies on a set of adjustable parameters. These parameters of PQCs are adjusted during QNN training.

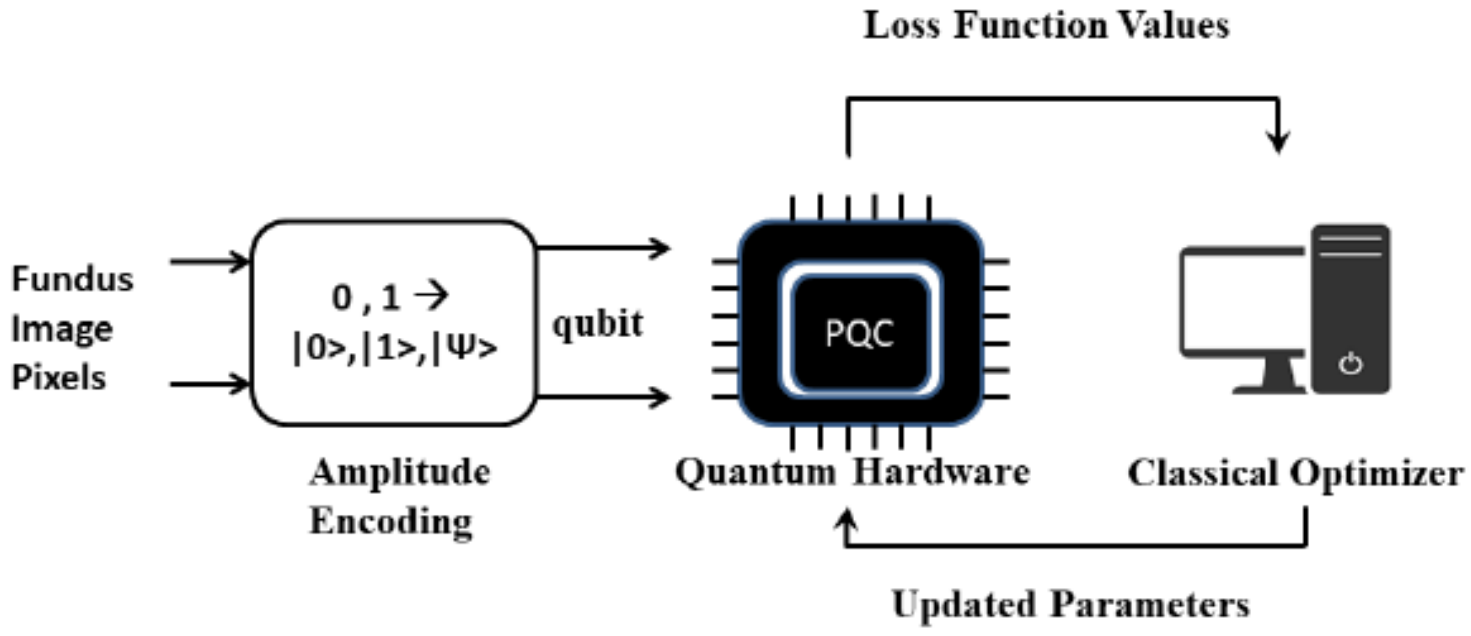

Fig. 3: Block diagram of Quantum Neural Network

## 4.4 Federated Learning Architecture

Federated learning is a machine learning algorithm, in which multiple institutions or nodes process their local data without sharing it with any other nodes or the server. This process enables nodes to keep their data secure. The local models are trained on local data and the model parameters are shared only to a federated server, where a global model is created by aggregating the local models. In our system, we used a federated weighted average for aggregation. If there are n local models with parameters $W_{i_l}$, and corresponding weight $a_{i_l}$, the global model uses equation 14 for computing federated weighted average. $W_{i_g}$ is the aggregated global weight.

$$W_{i_g} = \frac{\sum_{i=1}^{n} W_{i_l} * a_{i_l}}{\sum_{i=1}^{n} a_{i_l}} \tag{14}$$

The parameters w of the global model are shared with the local nodes. The local model with the new parameters learns from the local data again to improve accuracy. This process is repeated several times to improve the learning process.

### 4.5 Quantum Neural Network within Federated Learning Architecture

We used QNN within the federated learning framework. The algorithm 1 and the algorithm 2 represent local and global quantum federated learning algorithms, respectively.

**Algorithm 1** Distributed Quantum Federated Algorithm: Local QNN Learning

1: **Input**: Number of local nodes $n$, QNN model object at local nodes $W_i$, local iteration $i_l = 0$, learning rate $\eta$
2: **Output**: n number of local QNN models
3: **for** $i = 1$ to $n$ **do**
4: //Check for Local Convergence up to no improvement of loss.
5: **while** No improvement of *epsilon* **do**
6: //Training of local QNN model on local medical data at $i^{th}$ geographic location.
7: //Weight adjustment using Adam Optimizer:
8: $W_{i_{l+1}} = W_{i_l} - \eta \nabla \epsilon$
9: $W_i = W_{i_{l+1}}$
10: $i_l = i_l + 1$
11: Loss ($\epsilon$) computation.
12: **end while**
13: **end for**
14: **for** $i = 1$ to $n$ **do**
15: Local QNN model object $W_i$ is transmitted to aggregator cloud through wireless network.
16: **end for**

**Algorithm 2** Distributed Quantum Federated Algorithm: Global QNN Learning

1: **Input**: Number of local nodes $n$, Local QNN model object at global nodes $W_{i_l}$, global iteration $i_g = 0$, Global QNN model object at global nodes $W_{i_g}$, $a_{i_l}$ is the weight, learning rate $\eta$, Target accuracy $Acc$
2: **Output**: Trained aggregation model at local nodes
3: Weighted Average Aggregation $W_{i_g} = \frac{1}{\sum_{i=1}^{n} a_{i_l}} \sum_{i=1}^{n} W_{i_l} * a_{i_l}$
4: **for** $i = 1$ to $n$ **do**
5: Global QNN model object $W_{i_g}$ transmitted back to different local medical service providers' locations through the wireless network.
6: **end for**
7: **for** $i = 1$ to $n$ **do**
8: //Check for Global Convergence up to the target $Acc$ is achieved
9: **while** $!Acc$ **do**
10: //Training of returned back global QNN at $i^{th}$ local node.
11: Loss ($\epsilon$) computation.
12: //Global weight adjustment using Adam Optimizer:
13: $W_{i_g+1} = W_{i_g} - \eta \nabla \epsilon$
14: $W_{i_g} = W_{i_g+1}$
15: $i_g = i_g + 1$
16: **end while**
17: **end for**

## 5 Implementation and Experimental Results

The experiment is executed in Google Colaboratory and AWS Braket using PennyLane, a cross-platform Python library.

We created two models using the same number of MA-affected and MA-free image patches of size 7x6x3 from E-ophtha and Retina MNIST datasets. The models are validated using the images from the respective data set. In addition, we performed model cross-validation taking images from the Kaggle dataset.

### 5.1 Dataset Descriptions

**E-ophtha dataset [30] :** E-ophtha dataset has two subdatabases. E-optha MA subdatabase contains 148 images with MA or other small red lesions, 233 MA-free images. E-optha Exudate subdatabase contains 47 images with exudates, 37 images with no lesions. We executed experiments taking MA affected images as these images only carry the sign of early DR. The resolutions of E-ophtha images are $960 \times 1440$, $1000 \times 1504$, $1360 \times 2048$ and $1696 \times 2544$.
**Kaggle dataset [31]:** This is a large (88,702 images) dataset. The images are of high resolution (4752x3168). Clinicians graded the images for the presence of diabetic retinopathy as follows: 0 - No DR, 1- Mild DR, 2- Moderate DR, 3- Severe DR, 4- Proliferative DR. We executed experiments taking images affected by mild DR.
**Retina MNIST dataset [4] :** The images of the Retina MNIST data set are of 28x28 size. The number of images is 1600. The available image gradations are: 0-No DR, 1-Mild Non-Proliferative DR, 2-Moderate Non-Proliferative DR, 3-Severe Non-Proliferative DR, 4-Proliferative DR. We performed experiments taking images affected by mild Non-Proliferative DR.

The image patches are obtained from the database images and used for learning in federated learning framework. The details are presented in subsection 5.4.

### 5.2 Evaluation Metrics

Accuracy, precision, recall, F1-score and specificity are used as evaluation metrics for the FQPDR system.

$$Accuracy = \frac{TP + TN}{TP + TN + FP + FN} \tag{15}$$

$$Precision = \frac{TP}{TP + FP} \tag{16}$$

$$Recall = \frac{TP}{TP + FN} \tag{17}$$

$$F1 - score = \frac{2}{\frac{1}{Precision} + \frac{1}{Recall}} \tag{18}$$

$$Specificity = \frac{TN}{TN + FP} \tag{19}$$

where TP is the true positive, TN is the true negative, FP is the false positive and FN is the false negative. The F1 score is the harmonic mean of precision and recall, which shows how precise and robust the classifier is.

### 5.3 Parameterized Quantum Circuit

The performance of a parameterized quantum circuit (PQC) depends on various descriptors and its structure. Two main descriptors are expressibility and entangling capability. The expressibility of a PQC refers to its ability to generate diverse quantum states that well represent the Hilbert space. Expressibility typically saturates as the number of layers increases. Good expressibility also implies good entangling capability, as there is a strong correlation among qubits. Highly entangled and expressible quantum circuits are quite able to represent complex patterns, but they may also increase computational complexity and noise. We considered three different qubit topologies: nearest-neighbor (NN), circuit block (CB) and all-to-all (AA). The NN configuration uses a linear array of qubits, the CB configuration arranges qubits in a closed loop, and the AA configuration assumes a fully connected qubit graph. These configurations are illustrated in figure 4.

In the Nearest Neighbor circuit, the qubits interact with immediate neighbors. In a circuit-block configuration, qubits form blocks with interactions in a cyclic manner. In an all-to-all configuration, each qubit interacts with every other qubit. These configurations were evaluated for expressibility and entanglement capability by Sim, Sukin et al.[32]. The AA configuration showed the best expressibility and high entangling capability, though it requires more parameters, a greater circuit depth, and more complex qubit connectivity. The CB configuration has similar performance to AA but with slightly reduced complexity, while the NN configuration has low expressibility but requires the smallest circuit depth and circuit complexity.

Less expressible circuits may be cost-effective in circuit depth and gate count. Multilayering increases the expressibility of that circuit. There may be a circuit template that corresponds to a less favorable expressibility with limited layers but reaches a significant expressibility with sufficient layers. We propose an application-specific multilayer Nearest Neighbor circuit as a near-optimal choice. The circuit is shown in figure 5.

The proposed circuit is a real amplitude circuit which is used as a classification circuit in machine learning [29]. The circuit has alternating layers of Y rotations and Controlled-X entanglements with user-defined connection patterns. Prepared quantum states only have real amplitudes, and the complex part is always zero. The circuit parameter values change at the time of training by the optimizer. The qubit 1 is controlled by qubit 0. Similarly, qubit 2 is controlled by qubit 1. Following similar consecutive connections, the qubit 6 is controlled by all other qubits. Hence, the measurement is done only at qubit 6 to predict the existence of disease symptoms. In the suggested approach, the RY gate on each qubit performs rotation along the Y-axis. The angles of rotations are used as the parameters of the PQC. Entanglement is achieved through the application of the Controlled-X (CNOT) gate. A 7-qubit 2-layer PQC consisting of RY gates followed by Controlled-X gates of 2 layers with 14 parameters for 14 angles of RY gates is shown in figure 5.

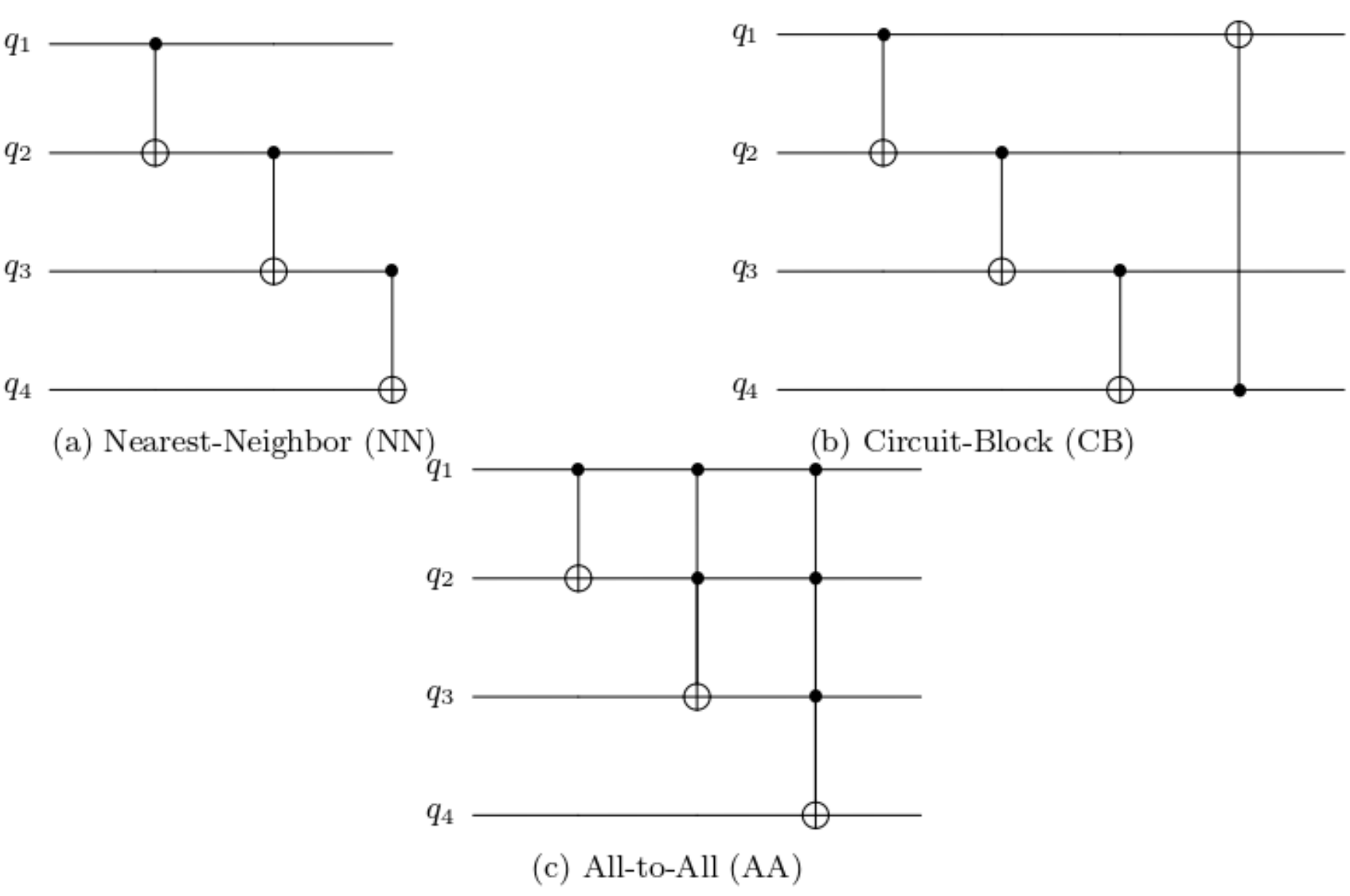

Fig. 4: Quantum Circuit Configurations: (a) Nearest-Neighbor (NN), (b) Circuit-Block (CB), and (c) All-to-All (AA)

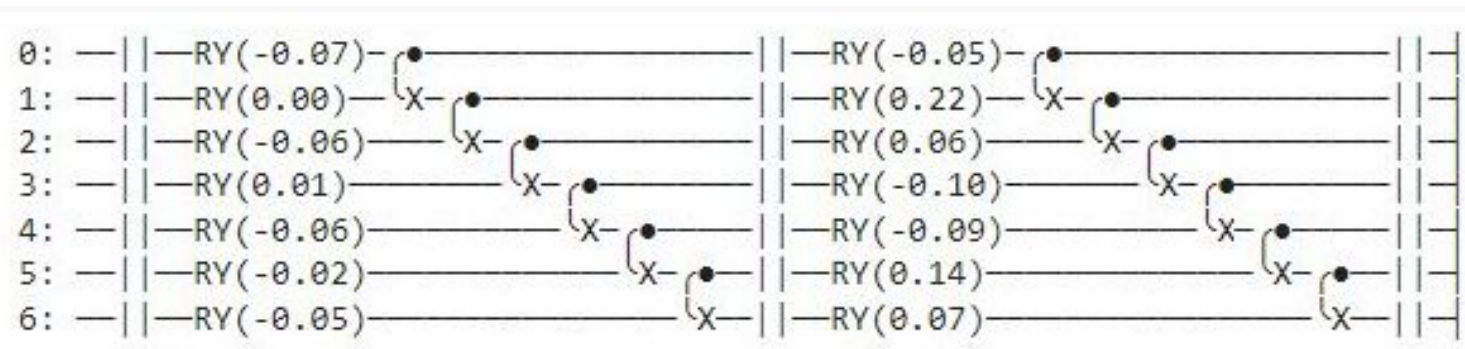

Fig. 5: Architecture of Parameterized Quantum Circuit

### 5.4 Training Set Preparation and Learning in Federated Framework

Centralized machine learning approach involves collecting and aggregating raw data in a repository where the machine learning model is trained. This approach incurs high storage and computational costs. Federated Learning (FL) is a recently introduced decentralized machine learning technique designed to address the problems of strict data privacy regulations and limited availability of data sets. It enables model training without transferring data, as the model is trained on various decentralized end devices containing the individual local data only. The central server receives updated weights from the local models. This approach ensures the privacy of data, a critical requirement in medical data diagnosis. We created three clients and positioned them on the client-ends, with the training data distributed among them. Each of the local models was trained with the corresponding client data. The weights were then aggregated into an array to form the global model. The global model was then tested on validation data for DR detection.

We extracted image patches of resolution 7x6x3 from the dataset images of variable sizes. These patches are fed to parameterized quantum circuit. Quantum neural networks at three virtual clients are trained with limited number of samples. A set of 314 image patches were divided into train:test ratio of 75%-25% for each client. Thus, 235 patches are used for model training in each of the client devices. We considered equal number of healthy and microaneurysm affected patches in the training set. Thus the problem of data imbalance has been sorted out. 79 test patches were used for evaluation purpose. 126 pixel intensity values are padded with two additional features so that features can be encoded following amplitude encoding scheme as discussed in the subsection 4.2. The encoded information is passed within the parameterized quantum circuit. In our experiment, the QNN runs on 3 different datasets. The QNN for each case converges for 3 different sets of parameters. At the central server the global model is created using weighted averaging approach where the local client model with better performance is having more weight than the less efficient local model as described in section 4.4 and following the equation 14. The weight distribution for the implementation of weighted average aggregation follows the ratio of 5:5:4 for 3 local clients, as the first two clients achieve better local accuracy compared to the other client.

We conducted our quantum FL experiments using different optimizers. The performances at three clients are evaluated with individual local models and the global federated model separately. Tables 5, 6 and 7 show the evaluation metrics for three different local nodes or clients using Gradient Descent, Nesterov Momentum and Adam optimizers respectively using both centralized approach and decentralized federated approach. Accuracy is higher for Federated QNN while training optimization takes place using Adam optimiser than Federated QNN by Nesterov Momentum optimiser and Gradient Descent optimiser.

Adaptive Moment Optimizer (Adam) is used for training the model. We have taken the maximum validation accuracy within 100 epochs as the conver-gence criteria of the model. Adam adapts the learning rate for each parameter individually, allowing parameters with larger gradients with smaller updates and smaller gradients with larger updates to minimize the loss value, which is responsible for increasing the accuracy. The learning is adjusted dynamically based on mean (first moment) and uncentered variance (second moment) of the gradients, for faster and stable convergence. In this optimization, the moving averages of the gradients and the square of the gradients are represented in equations (20) and (21) respectively.

$$v_t = \beta_1 * v_{t-1} + (1 - \beta_1) * g_t \tag{20}$$

$$s_t = \beta_2 * s_{t-1} + (1 - \beta_2) * g_t^2 \tag{21}$$

Here $g_t$ is the gradient at iteration t, $v_t$ is the moving average of the gradient in iteration t and $s_t$ is the moving average of the square of gradients at iteration t, $\beta_1$ and $\beta_2$ decay rates for first and second moments of the gradients. If $\alpha$ is the learning rate, $w_t$ and $b_t$ are the weight and bias at iteration t respectively, the weight and bias at iteration t+1, i.e. $w_{t+1}$ and $b_{t+1}$ are computed as in equations (22) and (23) respectively. The $\epsilon$ is a very small positive integer to avoid division by zero.

$$w_{t+1} = w_t - \alpha * \frac{v_t}{\sqrt{s_t + \epsilon}} \tag{22}$$

$$b_{t+1} = b_t - \alpha * \frac{v_t}{\sqrt{s_t + \epsilon}} \tag{23}$$

Table 5: Client wise QNN performances before and after applying Federated Learning (FL) for different local nodes using Gradient Descent optimizer

| **Local Node /Client** | **Validation Accuracy (%)** | | **Precision (%)** | | **Recall (%)** | | **F1-Score (%)** | | **Specificity (%)** | |
|---|---|---|---|---|---|---|---|---|---|---|
| | **Before FL** | **After FL** | **Before FL** | **After FL** | **Before FL** | **After FL** | **Before FL** | **After FL** | **Before FL** | **After FL** |
| **#1** | 84.62 | 83.33 | 90.32 | 92.86 | 75.68 | 70.27 | 84.36 | 82.86 | 92.68 | 95.12 |
| **#2** | 83.33 | 83.33 | 84.85 | 87.10 | 77.78 | 75.00 | 83.11 | 82.10 | 88.10 | 90.48 |
| **#3** | 81.01 | 82.28 | 88.46 | 88.89 | 65.71 | 68.57 | 79.97 | 81.42 | 93.18 | 93.18 |

### 5.5 FQPDR System Evaluation

Following quantum classical hybrid execution, both the centralized model and decentralized federated models are trained for 100 epochs, but the model with maximum validation accuracy is saved for transmission to the central server.

Table 6: Client wise QNN performances before and after applying Federated Learning (FL) for different local nodes using Nesterov Momentum optimizer

| **Local Node /Client** | **Validation Accuracy (%)** | | **Precision (%)** | | **Recall (%)** | | **F1-Score (%)** | | **Specificity (%)** | |
|---|---|---|---|---|---|---|---|---|---|---|
| | **Before FL** | **After FL** | **Before FL** | **After FL** | **Before FL** | **After FL** | **Before FL** | **After FL** | **Before FL** | **After FL** |
| **#1** | 85.97 | 85.90 | 93.33 | 93.33 | 75.67 | 75.67 | 85.61 | 85.61 | 95.12 | 95.12 |
| **#2** | 84.61 | 85.90 | 90.00 | 93.10 | 75.00 | 75.00 | 84.24 | 85.49 | 92.85 | 95.24 |
| **#3** | 81.01 | 86.07 | 88.46 | 85.29 | 65.71 | 82.85 | 79.98 | 85.55 | 93.18 | 88.64 |

Table 7: Client wise QNN performances before and after applying Federated Learning (FL) for different local nodes using Adam optimizer

| **Local Node /Client** | **Validation Accuracy (%)** | | **Precision (%)** | | **Recall (%)** | | **F1-Score (%)** | | **Specificity (%)** | |
|---|---|---|---|---|---|---|---|---|---|---|
| | **Before FL** | **After FL** | **Before FL** | **After FL** | **Before FL** | **After FL** | **Before FL** | **After FL** | **Before FL** | **After FL** |
| **#1** | 88.46 | 89.74 | 88.89 | 96.77 | 86.49 | 81.08 | 88.41 | 89.57 | 90.21 | 97.50 |
| **#2** | 88.46 | 88.46 | 93.54 | 90.91 | 80.55 | 83.33 | 88.23 | 88.31 | 95.23 | 92.85 |
| **#3** | 86.07 | 87.34 | 87.5 | 90.32 | 80.00 | 80.00 | 85.74 | 86.98 | 90.93 | 93.10 |

Table 8: Comparison between Non-FL and FL QNN training using Gradient Descent Optimizer

| Model | Dataset | Epoch | Exec time (sec) | Validation Accuracy |
|---|---|---|---|---|
| Non-FL QNN | E-ophtha | 51 | 524 | 84.62% |
| FL QNN | E-ophtha | 66 | 629 | 83.33% |

We present the comparison between federated and non-federated QNN training, in terms of the number of epochs, execution time in seconds, and validation accuracy with the Gradient descent optimizer, the Nesterov momentum optimizer, and the Adam optimizer and are shown in Tables 8, 9, and 10 respectively. It is notable that Federated QNN has better or comparable accuracy, precision, recall, F1-score, and specificity than Non-FL QNN for the E-ophtha dataset in most of the cases. The proposed method is applied to create FL-QNN models using E-ophtha and Retina MNIST images. These models are tested with 100 Kaggle dataset images for cross-validation of the FQPDR system. The cross-dataset evaluation performances to detect even the presence of mild DR are good and are shown in Table 11.

Table 9: Comparison between Non-FL and FL QNN training using Nesterov Momentum Optimizer

| Model | Dataset | Epoch | Exec time (sec) | Validation Accuracy |
|---|---|---|---|---|
| Non-FL QNN | E-ophtha | 69 | 699 | 85.9% |
| FL QNN | E-ophtha | 76 | 764 | 85.9% |

Table 10: Comparison between Non-FL and FL QNN training using Adam Optimizer

| Model | Dataset | Epoch | Exec time (sec) | Validation Accuracy |
|---|---|---|---|---|
| Non-FL QNN | E-ophtha | 40 | 271 | 88.46% |
| FL QNN | E-ophtha | 69 | 679 | 89.74% |

Table 11: Proposed Federated QNN based System Evaluation on Kaggle Dataset Images

| DR-labels by clinician | No of Kaggle images | Correctly identified by E-ophtha model | Correctly identified by Retina-MNIST model |
|---|---|---|---|
| 1 - Mild DR | 19 | 17 | 17 |
| 2 - Moderate DR | 19 | 17 | 14 |
| 3 - Severe DR | 19 | 19 | 19 |
| 4 - Proliferative DR | 19 | 19 | 19 |

### 5.6 Qualitative Assessment of the Proposed FQPDR system

By integrating quantum neural network and federated learning, the approach aims to achieve robust performance in classifying retinal images, ultimately contributing to improved diagnostic capabilities in ophthalmology.This is the rationale behind the selection of specific algorithms and techniques used in our system.

In the previous sections, we have included detailed case studies showing the classification of retinal image patches and the system's ability to identify affected patches. In addition, we incorporate a qualitative assessment of the test images by a medical professional. The test images are evaluated by a registered ophthalmologist from a renowned Govt. hospital, as well as by the FQPDR system model. It is notable that FQPDR is trained on the data set annotations itself and the system is quite capable to learn the annotations. The evaluation results are inspiring and shown in table 12.

Table 12: Qualitative Assessment of the Proposed FQPDR Model with Ophthalmologist Annotations

| **Sl No.** | **E-Ophtha Image Name** | **Dataset Annotation** | **Doctor Evaluation** | **FQPDR Evaluation** |
|---|---|---|---|---|
| 1 | C0000886 | NPDR | Normal | NPDR |
| 2 | C0000887 | NPDR | NPDR | NPDR |
| 3 | C0001629 | NPDR | Normal | NPDR |
| 4 | C0001883 | NPDR | NPDR | NPDR |
| 5 | C0001884 | NPDR | Normal | NPDR |
| 6 | C0001885 | NPDR | NPDR | NPDR |
| 7 | C0001886 | NPDR | Normal | NPDR |
| 8 | C0003222 | NPDR | Normal | NPDR |
| 9 | C0003226 | NPDR | NPDR | NPDR |
| 10 | C0004028 | NPDR | Normal | Normal |
| 11 | C0004029 | NPDR | Normal | NPDR |
| 12 | C0007331 | NPDR | Normal | NPDR |
| 13 | C0007333 | NPDR | NPDR | NPDR |
| 14 | C0008258 | NPDR | NPDR | NPDR |
| 15 | C0014790 | NPDR | Normal | NPDR |
| 16 | C0014793 | NPDR | NPDR | NPDR |
| 17 | C0015828 | Normal | Normal | Normal |
| 18 | C0015829 | NPDR | NPDR | NPDR |
| 19 | C0024411 | Normal | Normal | NPDR |
| 20 | C0024413 | Normal | NPDR | NPDR |
| 21 | C0024544 | NPDR | NPDR | NPDR |
| 22 | C0024545 | NPDR | NPDR | NPDR |
| 23 | DS0000DBO | NPDR | NPDR | NPDR |
| 24 | DS000DGS | NPDR | Normal | NPDR |
| 25 | DS000E6N | NPDR | NPDR | NPDR |
| 26 | DS000FGE | NPDR | NPDR | NPDR |
| 27 | DS000FGF | NPDR | Normal | Normal |
| 28 | DS000FMH | NPDR | Normal | NPDR |
| 29 | DS000FOK | NPDR | NPDR | NPDR |
| 30 | DS000FOL | NPDR | NPDR | NPDR |
| 31 | DS000FOM | NPDR | NPDR | NPDR |
| 32 | DS000IYR | Normal | Normal | Normal |
| 33 | DS000IYS | Normal | Normal | Normal |
| 34 | DS000IYV | Normal | Normal | Normal |
| 35 | DS000JHB | Normal | Normal | Normal |
| 36 | DS000JHC | Normal | Normal | Normal |
| 37 | DS000QA2 | NPDR | NPDR | NPDR |
| 38 | DS00008G | NPDR | NPDR | NPDR |
| 39 | DS00008I | NPDR | NPDR | NPDR |
| 40 | DS00008J | NPDR | NPDR | NPDR |

### 5.7 Comparison with Existing Works

We also present a comparative analysis, highlighting the strengths of our approach in relation to other state-of-the-art techniques. Comparisons with related literature are shown in Table 13. We created models from two datasets, highly different in resolutions. Retina MNIST images are 28x28x3, whereas E-ophtha images are 2544x1696x3. Both models are cross-evaluated with images from the Kaggle dataset. The existing classical methods usually require a lot of training parameters for converging to optimal accuracy. In table 13, the classical method [18] uses 968005 parameters to effectively learn the model. So, the method becomes computationally exhaustive, too. The proposed low power and light weight FQPDR uses only 15 parameters. With an increasing number of qubits, the computational power of a QPU increases exponentially, but power consumption does not follow a similar growth pattern. Classical computation power and power consumption both follow a linear growth pattern. Thus, QPU is considered to be low-power and computationally efficient compared to its classical counterpart. The loss and accuracy graphs of the proposed QNN and classical CNN are shown in figure 10a and figure 10b respectively for E-ophtha dataset. For comparison we considered maximum accuracies reported in [28] for QNN and orthogonal QNN. Number of learnable parameters, number of training samples with resolutions and comparative evaluation metric values are given in Table 13.

We have implemented a limited parameter classical CNN model which produces almost similar accuracy to the FQPDR model. This classical model applies a 3x3 convolution filter with a rectified linear unit activation function for feature extraction, followed by a 2x2 max-pooling layer for dimensionality reduction, and a flatten layer. A dense layer with a sigmoid activation function is used for affected or normal retinal image patch classification. The number of learnable parameters is 33. The number of epochs it requires to achieve the nearly similar accuracy is 62. We present QNN and CNN model comparison in Table 14.

## 6 Discussion

Medical Image Processing has been one of the main research areas in the last decade due to its importance to the well-being of mankind. Classifying retinal images falls under this area. In the mild DR stage, only microanurysms are present. More significant blood vessel blockages are present in moderate DR. In severe DR, more blood vessels are blocked. Proliferative is the advanced stage of DR, where the blood vessels are closed and new blood vessels begin to grow. The proposed FQPDR has emphasized training on microaneurysm images to address the problem of detecting mild DR because early detection of DR is the most challenging among all DR stages. The following are some of the points related to system analysis.

Table 13: Comparison of the Proposed Light-weight Federated QNN Model based DR Detection System with other existing DR detection methods.

| **Author-year** | **Learning Model/ Learnable Parameters** | **Objective** | **Database/ Number of Training Images and Resoluion** | **Pre-processing** | **Evaluation Accuracy** | **Precision** | **Recall** | **F1-Score** |
|---|---|---|---|---|---|---|---|---|
| Jagan Mohan N et al. 2023 [18] | Classical, FL **learnable parameters:** 968005 | A FL based approach to grade diabetic retinopathy from 0 to 4. | Messidor-2, IDRiD, Kaggle and local dataset, **train data:** 4000 high resolution fundus images | CLAHE, Entropy enhancement methods | 98.6% | 97.25% | - | 97.5% |
| Landman, Jonas, et al. 2022 [28] | Quantum assisted, Non-FL | Quantum assisted neural network to identify DR | Retina-MNIST dataset, **train data:** 1080 fundus images (28x28) | Dimension reduction using PCA | 80% | - | - | - |
| | Quantum, Non-FL **learnable parameters:**28 | Quantum orthogonal neural network to identify DR | | | 79% | - | - | - |
| Proposed Federated QNN | Quantum, FL **learnable parameters:** 15 | A FL and QNN based approach to identify DR | E-ophtha Dataset, **train data:** 234 fundus image patches (7x6x3) | Patch extraction from retinal images | 89.74% (AWS Braket) | 96.77% | 81.08% | 89.57% |
| | | To identify DR | Cross Validating with Kaggle Dataset | | 84.12% (AWS Braket) | 85.71% | 94.73% | 89.69% |
| | | To identify DR | Retina MNIST Dataset, train data: 234 fundus image patches (7x6x3) | | 75.61% (AWS Braket) | 75.68% | 71.94% | 75.48% |
| | | To identify DR | Cross Validating with Kaggle Dataset | | 81.0% (AWS Braket) | 85.19% | 90.79% | 88.11% |

1. **Use of QNN:** CNN is quite efficient in this job, but CNN requires a large number of learnable parameters to learn from huge amounts of data. The QNN circuit uses very few parameters to learn the same problem. We ex-perimented with a variety of PQCs for the early detection of retinopathy. But the circuit in figure 5 is quite efficient. QNN can learn characteristic correlations effectively from limited samples and with few learnable parameters compared to its classical counterparts, which is also observable in tables 13 and 14. QNN converges faster than classical CNN as we observe the number of epochs in the table 14. QNN outperforms the performance of CNN with comparable architectures.

Table 14: Comparison of the Proposed FQPDR Model with a limited parameter Classical CNN Model

| **Learning Model** | **Database** | **Model** | **No of Learnable Parameters** | **No of epochs** | **Accuracy** |
|---|---|---|---|---|---|
| FQPDR | E-ophtha dataset | Quantum Neural Network | 15 | 24 | 89.74% |
| Proposed limited parameter CNN | E-ophtha dataset | Classical Convolution Neural Network | 33 | 62 | 85.62% |

2. **Amplitude Encoding:** Amplitude encoding relies on wave function representation of quantum systems, where the amplitudes represent the data values. Amplitude encoding requires a logarithmic number of qubits to represent data sets having arbitrary numbers of samples and features, leading to a qubit-efficient learning procedure. Quantum efficiency and speedups are realizable with efficient state preparation.
3. **Use of Optimizer:** Different optimizers are also checked for optimization, but the Adam Optimizer performs the best among them. The optimizer runs for 70 epochs. The accuracy and loss update on each epoch is shown in figure 6a for the E-ophtha dataset. The accuracy and loss update for 70 epochs using the Nesterov Momentum Optimizer is shown in figure 6b , the same using Gradient Descent Optimizer is shown in figure6c. Gradient Descent (GD) maintains a single learning rate for all weight updates and the learning rate does not change during model training. GD is not subject to variance since at each step the average gradient is computed using the whole data set. The downside is that every step becomes computationally expensive, O(nd) per iteration, where n is the number of samples in our dataset and d is the number of dimensions of the data. But Adam (Adaptive Moment Estimation) is a better choice as it is an adaptive optimization algorithm that maintains a separate learning rate for each network weight and adapts it as learning unfolds. It uses the running average of the gradients, giving better convergence over GD. But Adam involves additional computation costs for maintaining moving averages, as well as bias correction terms. Nesterov Momentum or Nesterov's Accelerated Gradient

Descent works by following the negative gradient of an objective function for convergence and the convergence of gradient descent accelerates by adding Nesterov Momentum. The algorithm works by updating the weight using the previous weight update and the current gradient concerning the parameters. It often manifests faster convergence compared to GD, but has a bit higher computational cost considering the momentum. Train and validation accuracy curves with the number of epochs for Non-FL and FL versions of the E-ophtha data set using Adam optimizer are shown in figure 7a and figure 7b.

4. **Federated QNN Architecture:** The number of Epoch vs NonFL-Accuracy and number of Epoch vs NonFL-Loss using the E-ophtha dataset are shown in figure 8a, figure 8b respectively. The number of Epochs vs. FL-Accuracy and number of Epochs vs. FL-Loss using the E-ophtha dataset are shown in figures 8c and 8d, respectively. Tables 8, 9, and 10 show the performances of the FL and non-FL versions using three optimizers. The centralized approach may not fully capture the diversity of data distributed across various nodes. Furthermore, the federated learning approach preserves the privacy of medical information because the confidential information is not shared with the server. The PQC parameters are sent to the server, and aggregated values of the parameters are returned to the clients. The parameters, i.e., the weights and bias of the PQC, are updated on each epoch to obtain optimal accuracy. The variation of non-FL weights, non-FL bias, FL weights and FL bias with each epoch using the E-ophtha dataset are shown in figure 9a, figure 9b, figure 9c and figure 9d, respectively. These four plots indicate how all 15 parameters (14 weights + 1 bias) change during training convergence. For nonFL weights, all weights start at the same zero value as shown in 9a. They attain different values during convergence. After global aggregation, all the weights have different values. Hence, FL weights start from different values as shown in 9c. In FL, local data training can be performed on local nodes in parallel, increasing the overall speed of the system.
5. **QNN vs. QCNN** The quantum convolutional neural network (QCNN) uses a quantum convolution layer and a quantum pooling layer. We have tested the experiment with a QCNN where the quantum convolution layer consists of RY and CX gate and the pooling layer consists of CRZ and CRX gate, with the total number of 1-qubit gates being 20 and the total number of 2-qubit gates being 28. We have optimized the number of gates of the circuit for maximum validation accuracy, and ultimately get a QNN with 14 1-qubit gates and 12 2-qubit gates as shown in figure 5. Comparisons of the validation accuracy of our Non-FL QNN with this Non-FL QCNN using Adam optimizer are shown in Table 15. Although with increasing number of parameters the expressive power is increased, QNN also becomes susceptible to noise. The declination of QCNN could be attributed to the increased complexity and noise susceptibility of the circuit. As quantum computing hardware becomes more accessible, with greater availability of noise-free qubits, better learning can be done in the future.

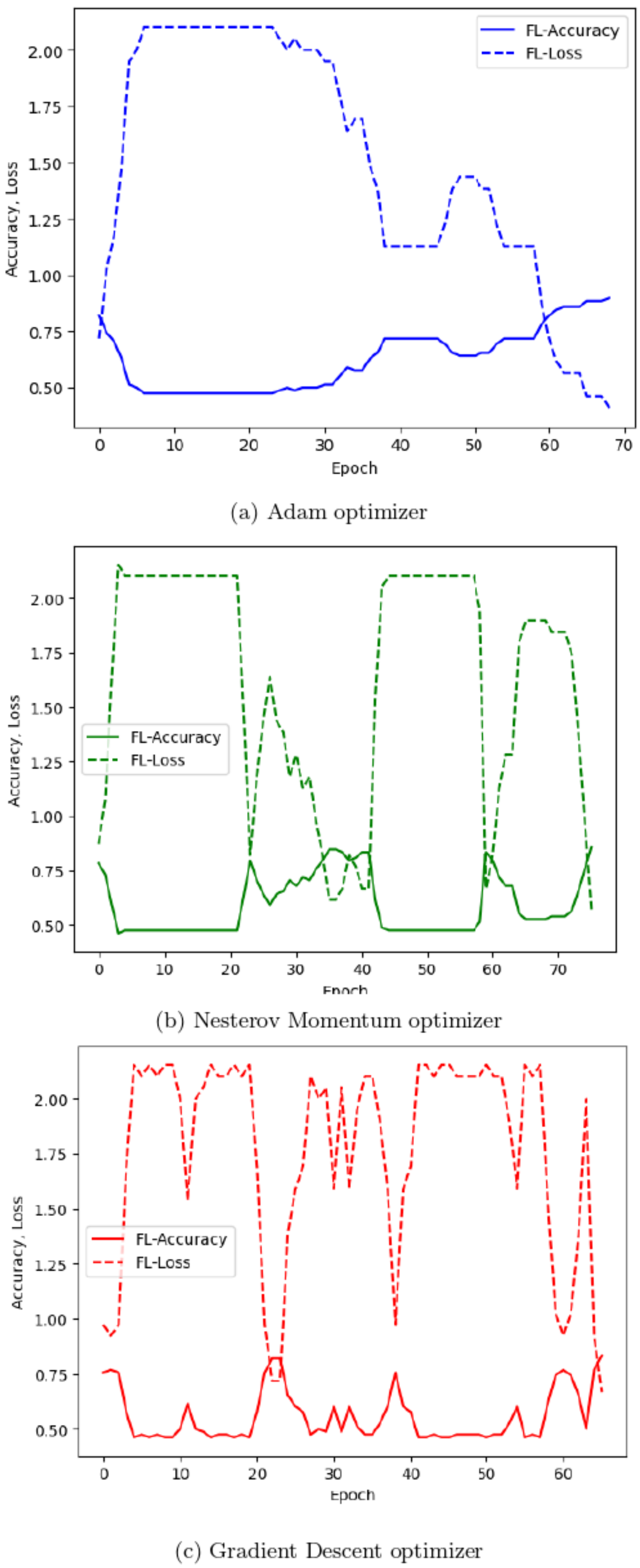

(a) Adam optimizer

(b) Nesterov Momentum optimizer

(c) Gradient Descent optimizer

Fig. 6: Epoch vs FL-Accuracy, FL-Loss curves of E-ophtha Model using three different optimizers

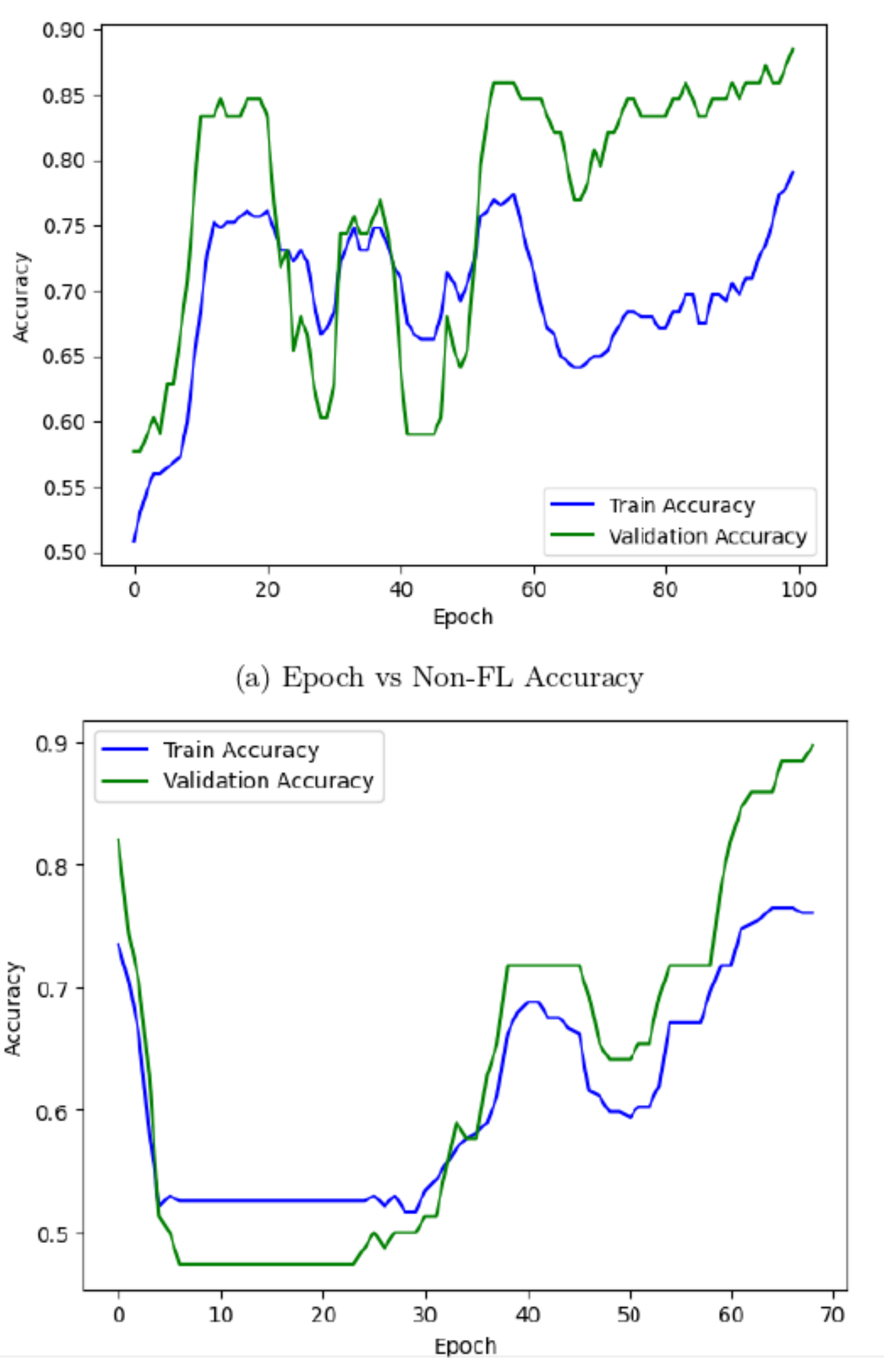

(a) Epoch vs Non-FL Accuracy

(b) Epoch vs FL Accuracy

Fig. 7: Epoch vs Training Accuracy, Validation Accuracy curves of Non-FL and FL QNN for E-Ophtha model using Adam Optimizer

Table 15: Comparison of validation accuracy between Non-FL QNN and Non-FL QCNN using Adam optimizer

| Circuit | Datasets | Local Model 1 | Local Model 2 | Local Model 3 |
|---|---|---|---|---|
| Non-FL QNN | E-ophtha | 88.46% | 88.46% | 86.08% |
| Non-FL QCNN | E-ophtha | 85.90% | 88.46% | 83.54% |

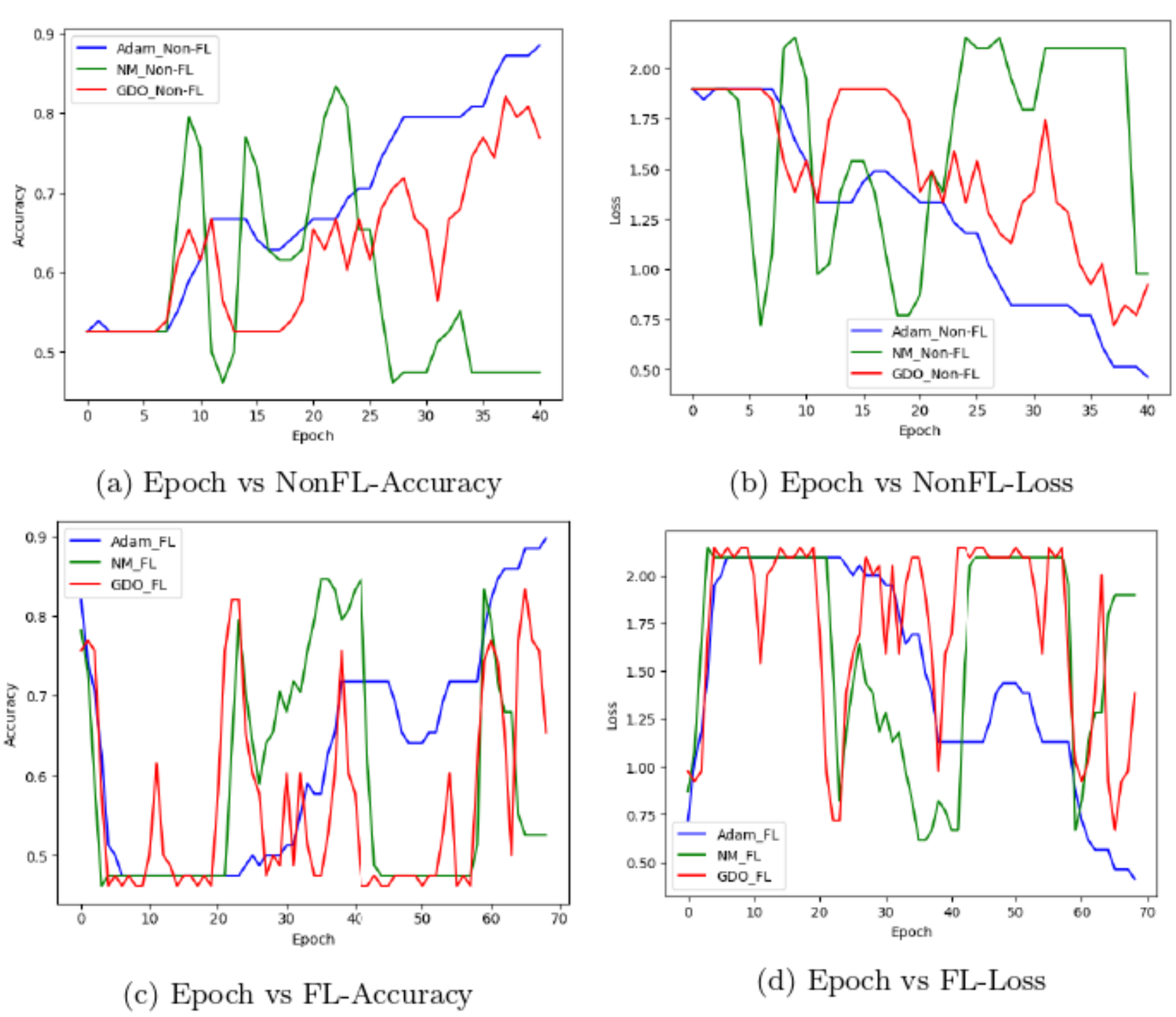

(a) Epoch vs NonFL-Accuracy

(b) Epoch vs NonFL-Loss

(c) Epoch vs FL-Accuracy

(d) Epoch vs FL-Loss

Fig. 8: Epoch vs NonFL-Accuracy, NonFL-Loss, FL-Accuracy, FL-Loss using Adam, Nesterov Momentum and Gradient Descent Optimizers for E-Ophtha Model

The Internet facilitates the connection of diverse edge devices, enabling the establishment of remote healthcare systems through the Internet of Medical Things (IoMT), connecting patients with healthcare professionals. Smart phones/tablets/laptops would be used to capture images of the retina. These edge devices would train the local models and send the parameters to the federated server. Especially, FQPDR could encourage new explorations in the domain of secure quantum machine learning based IoMT applications.

## 7 Potential Limitations

QNN applications resolved certain problems more efficiently than the classical methods. It has the potential to revolutionize machine learning by improving network training with few learnable parameters and limited samples. However, it poses some significant challenges that need to be addressed.

– Quantum hardware resources are currently expensive and not widely accessible.

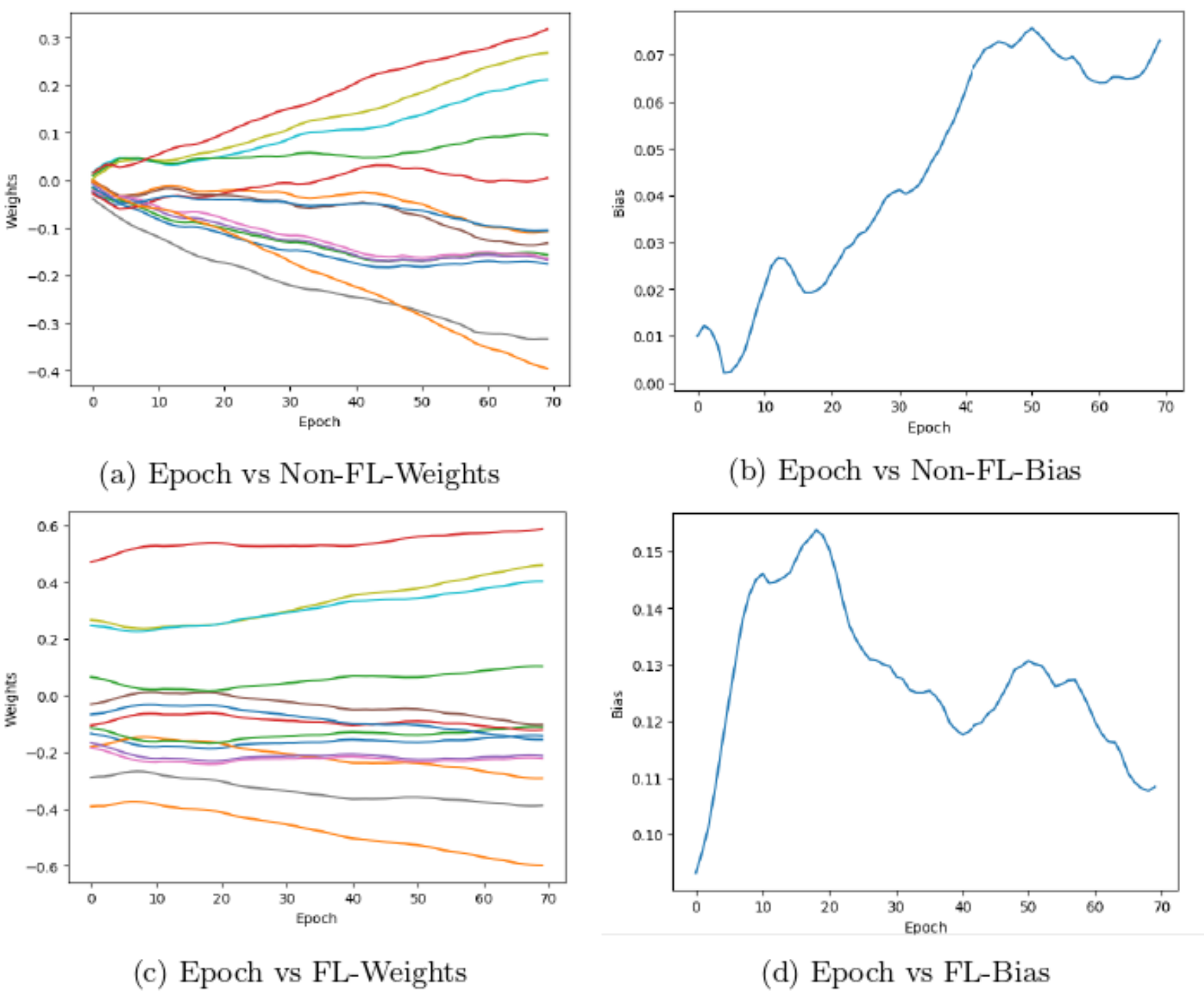

(a) Epoch vs Non-FL-Weights

(b) Epoch vs Non-FL-Bias

(c) Epoch vs FL-Weights

(d) Epoch vs FL-Bias

Fig. 9: Epoch vs Weights, Epoch vs Bias for Non-FL and FL Training of E-Ophtha Model using Adam optimiser

- Contemporary quantum computers have a limited number of qubits, and these qubits suffer from errors due to decoherence and noise.
- Hybrid QNN training encompasses both classical and quantum execution part, which is still an open research direction for implementing real-time diagnostic system.
- Quantum encoding and feature mapping may produce an overhead to the application

Despite its limitations, the presented model establishes a foundational approach to retinal image classification using federated QNN. Future improvements could adapt the network for multi-class classification, enabling it to identify specific types of abnormalities, such as different grades of DR. Advances in quantum hardware, improved quantum algorithms, and better understanding of quantum systems help overcome these limitations.

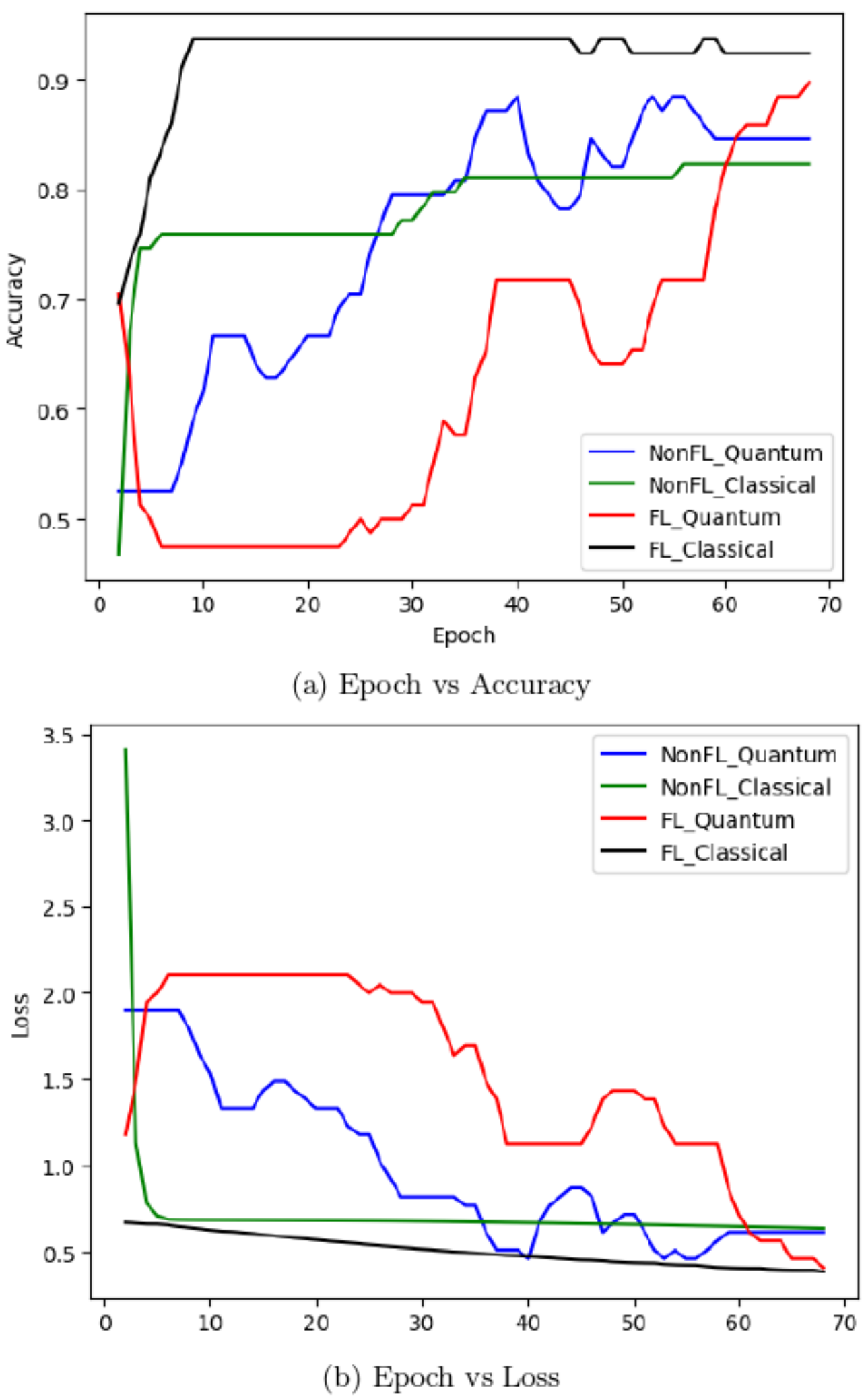

(a) Epoch vs Accuracy

(b) Epoch vs Loss

Fig. 10: Epoch vs Accuracy, Loss for Non-FL/FL classical and quantum neural networks using E-Ophtha Model

## 8 Conclusion

In the proposed FQPDR system, DR-affected retinal images are identified in the early stages using Federated QNN. Most of the previous methods used classical CNN for this problem. As quantum computing is rapidly developing and being applied in different fields for better feature representation with less number of samples and faster computations, we have designed a QNN for early detection of DR. Experimentally, it is verified that Adam optimizer produces the best accuracy. The proposed Federated QNN architecture preserves the privacy of the dataset at the edge devices. The evaluation metrics for accuracy, precision, recall, F1 score, and specificity are also encouraging.

In future, we intend to design and experiment with different types of parameterised quantum circuits for the problem to improve the system efficiency. The system can also be used for other medical images in an application specific way.

## Acknowledgement

We would like to acknowledge the Meity QCAL project, Ministry of Electronics and Information Technology, Government of India for the research grant and AWS for the Amazon Braket execution interface.

We further would like to acknowledge Dr. Anindya Gupta, Associate Professor, Regional Institute Of Ophthalmology, Medical College, Kolkata, India for his cooperation and support in evaluating the proposed FQPDR system qualitatively.

## Declarations

– Funding : This work is funded by Meity QCal, Ministry of electronics and Information Technology, Govt. of India and is supported by AWS for the Braket interface they provide to execute the programs.
– Competing interests : There are no competing interests.